\documentclass[10pt]{iopart} 
\usepackage{epsf,iopams,cite,graphicx,psfrag}
\newcommand{\ds}{\displaystyle}
\newcommand{\reminder}[1]{}
\begin{document}

%
\title{Classification of integrable clusters of classical Heisenberg
       spins}
\author{Marco Ameduri$^{\dag}$, Bogomil Gerganov$^{\dag}$ and Richard
        A Klemm$^{\ddag}$}
\address{$^{\dag}$Weill Cornell Medical College in Qatar,
        Education City, \makebox{P.\ O.\ Box 24144}, Doha, Qatar}
\address{$^{\ddag}$Department of Physics, Kansas State University,
        Manhattan, KS 66506, USA}
\eads{\mailto{ma22@cornell.edu}, \mailto{beg2@cornell.edu},
        \mailto{klemm@phys.ksu.edu}}
\begin{abstract}
In this letter we present a general classification of integrable
models of identical classical spins coupled via the isotropic
Heisenberg Hamiltonian. Our constructive proof of integrability
provides a solution scheme for the equations describing the time
evolution of each spin. We show that for a broad class of Heisenberg
couplings, it is possible to solve the equations of motion iteratively
by building a hierarchy of spins, each of which precesses about the
composite spin of the previous level. This leads to the identification
of the frequencies of precession at each level.
\end{abstract}



\section{Introduction}

The study of the dynamics of classical Heisenberg spins has long
proved invaluable in a variety of physical contexts. From a
mathematical point of view, spin clusters have been a natural testing
ground to explore the emergence of chaos \cite{sriva1}. The recent
developments in the experimental study of magnetic molecules
\cite{Cr3tetra:2,Fe3tetra:2,Mn12} have generated a renewed interest in
the dynamics of classical Heisenberg spins. Integral representations
of the time-dependent spin-spin correlation functions have been
presented for a variety of small integrable clusters
\cite{dimer,square,isosceles,dimer:field}. Analogous results for the
equivalent neighbor model or for an integrable deformation of it have
also been obtained \cite{eq_neighbour,squashed}. Among the more recent
developments, in \cite{knoxville} the neutron scattering structure
factors for a wide range of finite quantum spin systems were computed
analytically by Haraldsen, Barnes, and Musfeldt.

In this letter we report on a broad classification of integrable
cases. In section \ref{models} we present a class of models leading to
integrable equations of motion. In section \ref{solutions} the
integrability of the models is proved and the precession frequencies
are obtained. Section \ref{conclusions} contains our conclusions.


\section{Classification of the integrable  clusters}
\label{models}

The most general Heisenberg Hamiltonian for $N$ classical spins
normalized according to $|\bi{S}_{i}|=1$, $i=1, \ldots, N$, is
\begin{equation}
  H = \frac{1}{2} \sum_{i,j=1}^{N} g_{ij} \bi{S}_{i} \cdot \bi{S}_{j}
    = \sum_{i<j}^{} g_{ij} \bi{S}_i \cdot \bi{S}_j\;.
  \label{Ham:gen}
\end{equation}
The matrix of exchange couplings $\left[ g_{ij} \right]$ in
(\ref{Ham:gen}) is symmetric, $g_{ij}=g_{ji}$, and $g_{ii}=0$.
The time evolution of each spin is determined by the differential
equations
\begin{equation}
  \dot{{\bi{S}}}_i = \sum_{j=1}^{N} g_{ij} \bi{S}_i
                     \times \bi{S}_j \; .
  \label{EoM:gen}
\end{equation}
The structure of the equations of motion (\ref{EoM:gen}) explicitly
shows that the norm of each spin is indeed constant. The equations
(\ref{EoM:gen}) are non-linear and, for an arbitrary matrix of
couplings $\mathbf{G} =\left[ g_{ij} \right]$, they do not admit a
general solution in closed form. In the following we identify a class
of coupling matrices for which the general solution can be explicitly
constructed.

Let us partition the set $I = \left\{1,2,3,\ldots,N\right\}$ of spin
labels into nonintersecting subsets $A$, $B$, $C$, \dots, such that
\begin{equation}
  A \cup B \cup C \cup \ldots = I \; ,
  \qquad A \cap B = \emptyset \, , ~~ 
         A \cap C = \emptyset \, , ~ \ldots
  \label{part}
\end{equation}
We are going to call the set of spins whose labels are in $A$ the
`spin cluster $A$,' and so on for every other subset of $I$.  It is
useful to introduce also the following terminology.  A coupling
$g_{ij}$ is called internal if its indices $i$ and $j$ belong to the
same cluster, whereas a coupling $g_{ij}$ is called external if $i$
and $j$ belong to different clusters.

As we shall prove in section \ref{solutions}, if there is a partition
(\ref{part}) of $I$, for which all the external couplings are equal,
the problem of solving the equations of motion (\ref{EoM:gen}) is
reduced to separately solving the spin systems described by the index
sets $A$, $B$, etc.  If all the external couplings are equal to $J$,
the coupling matrix $\mathbf{G}$ has the form
\begin{equation}
  \mathbf{G} = \left(
               \begin{array}{cccccc}
                 \mathbf{G}^A & J\mathbf{E}  & J\mathbf{E}
                 & ~~~~~& ~~~~~& ~ \\[4pt]
                 J\mathbf{E} & \mathbf{G}^B & J\mathbf{E}
                 & ~~~~~ & ~~~~~ & ~ \\[4pt]
                 J\mathbf{E}    & J\mathbf{E}   & \mathbf{G}^C
                 & ~~~~~& ~~~~~& ~\\[4pt]
                 ~&~&~& \bullet &~&~ \\
                 ~&~&~&~& \bullet &~\\
                 ~&~&~&~&~& \bullet
      \end{array}
    \right) \, ,
  \label{Red:Matrix}
\end{equation}
where $\mathbf{G}^A$ is the matrix of the internal couplings between
spins in cluster $A$, $\mathbf{G}^B$ the matrix of the internal
couplings between spins in cluster $B$, and so on.  $\mathbf{E}$
is a matrix of appropriate dimension, all elements of which are equal
to $1$.

Moreover, we claim that if each of the coupling matrices
$\mathbf{G}^A$, $\mathbf{G}^B$, \ldots, can in turn be written in the
form (\ref{Red:Matrix}), with a single coupling $J_A$ among the
sub-clusters of $A$, a single coupling $J_B$ among the sub-clusters
of $B$, etc., ($J_A$, $J_B$, \ldots, need not be equal to $J$), and if
this can be carried on until we get clusters containing only single
spins, the equations of motion (\ref{EoM:gen}) can be solved
exactly. As an example, an 8-spin model of this type is
shown in figure 1. We also notice that all the models studied in
\cite{dimer,square,isosceles,dimer:field,eq_neighbour,squashed} fit
into this scheme.


\begin{figure}[htb]
\psfrag{1}{$1$}
\psfrag{2}{$2$}
\psfrag{3}{$3$}
\psfrag{4}{$4$}
\psfrag{5}{$5$}
\psfrag{6}{$6$}
\psfrag{7}{$\hspace{-1.5mm}7$}
\psfrag{8}{$\hspace{-1.5mm}8$}
\psfrag{A}{$J_1$}
\psfrag{B}{$J_2$}
\psfrag{C}{$J_4$}
\psfrag{D}{$J_3$}
\hspace{25mm}
\includegraphics[width=10cm]{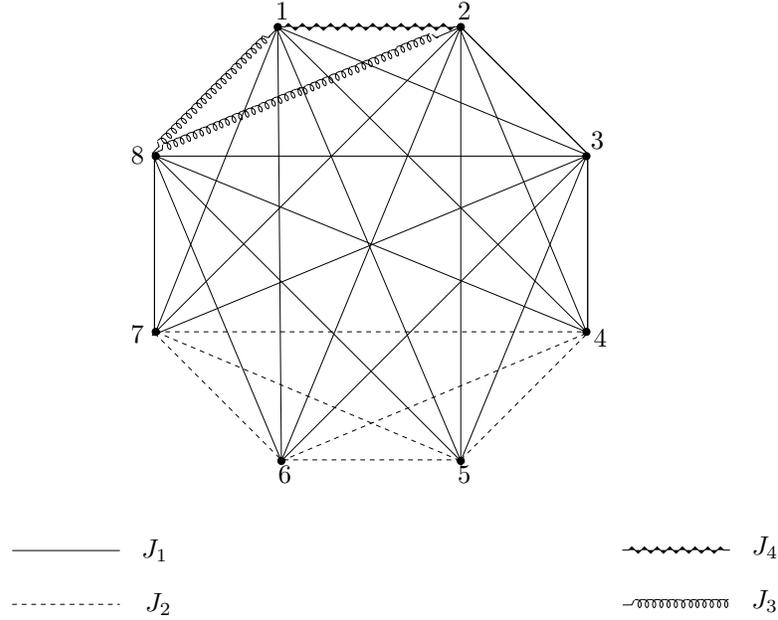}
\caption{A model of 8 interacting spins with 4 different
  couplings. Spins $\bi{S}_1$ and $\bi{S}_2$ form a dimer with a
  coupling $J_4$. Spin $\bi{S}_8$ is coupled to this dimer with a
  coupling $J_3$ to form an isosceles triangle. Spins $\bi{S}_4$,
  $\bi{S}_5$, $\bi{S}_6$, and $\bi{S}_7$ form a tetrahedron with
  coupling $J_2$. These individual clusters, together with the single
  spin $\bi{S}_3$, are coupled to one another by the external coupling
  $J_1$.}
\end{figure}


\section{Solution method and precession frequencies}
\label{solutions}

If the coupling matrix has the form (\ref{Red:Matrix}), the equation
of motion for a spin belonging to cluster $A$ is
\begin{equation}
        \label{EoM:redA}
  \dot{\bi{S}}_{\alpha} = \sum_{\alpha' \in A} g^{A}_{\alpha\alpha'}
                               \bi{S}_{\alpha} \times \bi{S}_{\alpha'}
                               + J\bi{S}_{\alpha} \times
                               \bi{S}_{\bar{A}}\,, \quad \forall
                               \alpha \in A
\end{equation}
where $\ds \bi{S}_{\bar{A}} \equiv \sum_{i \in I \setminus A}
\bi{S}_i$ is the sum of all the spins outside cluster $A$. Similar
equations can be written for the spins in every other cluster $B$,
$C$, etc.

With the definitions
\begin{equation}
  \bi{S}_{A} = \sum_{\alpha \in A} \bi{S}_{\alpha} \, ,
  \quad
  \bi{S}_{B} = \sum_{\beta \in B} \bi{S}_{\beta} \, ,
  \quad \ldots
  \label{spin:sums}
\end{equation}
we obtain the equations of motion describing the time evolution of the
spin sums (\ref{spin:sums}) by adding the equations (\ref{EoM:redA})
for all spins in a given cluster. E.g., for cluster $A$ we obtain
\begin{equation}
  \dot{\bi{S}}_A = J\bi{S}_A \times \bi{S}_{\bar{A}} \, .
  \label{A:sum} 
\end{equation}
Since the total spin $\bi{S}$ of the system can be written as
\begin{equation}
  \bi{S} = \bi{S}_A + \bi{S}_{\bar{A}}
  \label{S:tot}
\end{equation}
equation (\ref{A:sum}) becomes
\begin{equation}
  \dot{\bi{S}}_{A} = J \bi{S}_{A} \times \bi{S} \, .
  \label{A:prec}
\end{equation}

Since the total spin $\bi{S}$ is conserved in the Heisenberg model,
equation (\ref{A:prec}) can be solved immediately. The solution
describes the precession of the vector $\bi{S}_{A}$ around
$\bi{S}$. Choosing a coordinate system such that $\bi{S} = S
\mathbf{\hat{z}}$, the solution can be written as
\begin{equation}
\label{A:sol}
        \bi{S}_A(t) = S_{A}^{\parallel} \mathbf{\hat{z}}
        + S_{A}^{\perp} \left[\mathbf{\hat{x}}\cos(\omega t - \phi_A)
        - \mathbf{\hat{y}}\sin(\omega t - \phi_A)\right]
\end{equation}
where $\omega=JS$, $S_{A}^{\parallel}$ is the projection of $\bi{S}_A$
along the direction of $\bi{S}$, $S_{A}^{\perp}$ is its projection
perpendicular to the direction of $\bi{S}$, and $\phi_A$ is the initial
phase of $\bi{S}_{A}$.  Since all the $\bi{S}_A(t)$ and
$\bi{S}_{\bar{A}}(t)=\bi{S}-\bi{S}_{A}(t)$ are now explicitly known
functions of time, solving equation (\ref{EoM:redA}) is reduced to
solving the internal coupling sector for each cluster. Indeed, let us
focus again on cluster $A$ and let us assume that its index set $A$
can itself be partitioned in a way similar to (\ref{part}), i.e.,
\begin{equation}
        A = M \cup N \cup P \cup \ldots\;,
        \qquad M \cap N = \emptyset\,, ~~ M \cap P = \emptyset\,,
        ~ \ldots\;,
\label{part:A}
\end{equation}
\reminder{part:A}
and that the coupling matrix $\mathbf{G}^A$ has the structure
(\ref{Red:Matrix}), with all external couplings between the
sub-clusters $M$, $N$, $\ldots$ , being equal to $J_A$. The equations
of motion involving the spins of sub-cluster $M$ can then be written
as
\begin{equation}
        \label{EoM:redM}
        \dot{\bi{S}}_{\mu} = \sum_{\mu' \in M}
        g^M_{\mu\mu'} \bi{S}_{\mu} \times \bi{S}_{\mu'} +
        J_A\bi{S}_{\mu} \times \bi{S}_{\bar{M}} + J\bi{S}_{\mu} \times
        \bi{S}_{\bar{A}}\,, \; \forall \mu \in M
\end{equation}
where $\ds \bi{S}_{\bar{M}} \equiv \sum_{i \in A \setminus M}
\bi{S}_i$.  Once again, equations of the same form will hold for the
spins in the other sub-clusters of $A$.

Adding all the equations for the spins in sub-cluster $M$ in a way
similar to the derivations of (\ref{A:sum}) and (\ref{A:prec}), and
defining $\ds \bi{S}_{M} = \sum_{\mu \in M} \bi{S}_{\mu}$, we obtain
\begin{equation}
        \dot{\bi{S}}_M =  J_A\bi{S}_M \times \bi{S}_A
        + J\bi{S}_M \times \bi{S}_{\bar{A}}
\label{M:sum}
\end{equation}
\reminder{M:sum}
which can also be written in the form
\begin{equation}
        \dot{\bi{S}}_M = J\bi{S}_M \times \bi{S} + (J_A
        - J)\bi{S}_M \times \bi{S}_A\;.
\label{2-level:M}
\end{equation}

As long as we can apply the reduction scheme of Section 2 to the next
level of sub-clusters, this solving procedure can be carried on
iteratively. For the purpose of illustration, suppose that the
internal couplings between the spins in sub-cluster $M$ are all equal
to $J_M$, i.e., that the reduction procedure ends at this
level.\footnote{%
  This is the same as saying that the sub-clusters of $M$ consist of
  individual spins, each coupled to all the others with coupling
  $J_M$, or that $M$ is an equivalent neighbor model with respect to
  its internal couplings.}
The equations of motion for the individual spins in cluster $M$ can
then be written as
\begin{equation}
\hspace{-10mm}
\label{3-level}
        \dot{\bi{S}}_i = J\bi{S}_i \times \bi{S} + (J_A
        - J)\bi{S}_i \times\bi{S}_A + (J_M - J_A)\bi{S}_i \times
        \bi{S}_M, \; i \in M.  
\end{equation}
In order to write the solution for $\bi{S}_i(t)$ in a closed form we
apply successively the Fourier transform method used in
\cite{square,isosceles, squashed} to find $\bi{S}_A(t)$, then
$\bi{S}_M(t)$, and finally $\bi{S}_i(t)$, $i \in M$. Though cumbersome
and tedious, this is straightforward and mechanical. For the purposes
of this letter, we restrict ourselves to stating the resulting
frequencies of the various spin precessions : $JS$, $(J_A-J)S_A$, and
$(J_M-J_A)S_M$. Figure 2 illustrates the precessions of the
individual spins and the spin composites for the 8-spin model in
figure 1.


\begin{figure}[htb]
\psfrag{1}{$\bi{S}_1$}
\psfrag{2}{$\bi{S}_2$}
\psfrag{3}{$\bi{S}_3$}
\psfrag{4}{$\bi{S}_4$}
\psfrag{5}{$\bi{S}_5$}
\psfrag{6}{$\bi{S}_6$}
\psfrag{7}{$\bi{S}_7$}
\psfrag{8}{$\bi{S}_8$}
\psfrag{A}{$\bi{S}_{4567}$}
\psfrag{B}{$\bi{S}_{12}$}
\psfrag{C}{$\bi{S}_{128}$}
\psfrag{S}{$\bi{S}$}
\hspace{25mm}
\includegraphics[width=10cm]{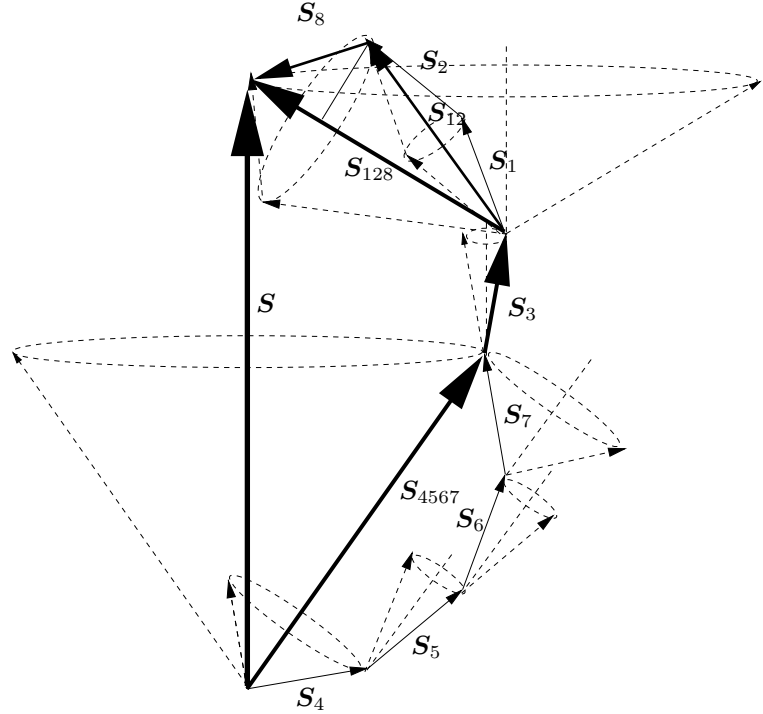}
\caption{The time evolution of the 8-spin model of figure 1.  The net
  spin of the isosceles triangle, $\bi{S}_{128}$, the net spin of the
  tetrahedron, $\bi{S}_{4567}$, and the individual spin $\bi{S}_3$
  precess with angular speed $J_{1}S$ about the total spin of the
  system, $\bi{S}$. The individual spins $\bi{S}_4$, $\bi{S}_5$,
  $\bi{S}_6$, and $\bi{S}_7$ in turns precess about
  $\bi{S}_{4567}$. The dimer spin $\bi{S}_{12}$ and the individual
  spin $\bi{S}_8$ precess about the isosceles spin $\bi{S}_{128}$.
  Finally, $\bi{S}_1$ and $\bi{S}_2$ precess about $\bi{S}_{12}$.}
\end{figure}

        More generally, if the reduction scheme leads to a hierarchy
of $n$ levels of sub-clusters, $i \in I^{(n)} \subset I^{(n-1)}
\subset \ldots \subset I^{(2)} \subset I^{(1)} \subset I$, where each
$k$-level cluster has a coupling $J_{(k)}$ among its $(k+1)$-level
sub-clusters, the time evolution of the $i^{\rm th}$ spin $\bi{S}_i$
can be found by iteratively solving the equations
\begin{eqnarray}
\hspace{-20mm}
\nonumber
        \dot{\bi{S}} & = 0
\\
\nonumber
\hspace{-20mm}
        \dot{\bi{S}}_{(1)} & = J\bi{S}_{(1)} \times \bi{S}
\\
\label{level:n}
\hspace{-20mm}
        \dot{\bi{S}}_{(2)} & = J\bi{S}_{(2)} \times \bi{S}
        + (J_{(1)} - J)\bi{S}_{(2)} \times \bi{S}_{(1)}
\\
\hspace{-20mm}
\nonumber
\ldots &
\\
\nonumber
\hspace{-20mm}
        \dot{\bi{S}}_{(n)} & = J\bi{S}_{(n)} \times \bi{S}
        + (J_{(1)} - J)\bi{S}_{(n)} \times \bi{S}_{(1)}
        + \ldots + (J_{(n-1)} - J_{(n-2)})\bi{S}_{(n)}
        \times \bi{S}_{(n-1)}
\\
\nonumber
\hspace{-20mm}
        \dot{\bi{S}}_i & = J\bi{S}_i \times \bi{S}
        + (J_{(1)} - J)\bi{S}_i \times \bi{S}_{(1)} + \ldots
        + (J_{(n)} - J_{(n-1)})\bi{S}_i \times \bi{S}_{(n)}
\end{eqnarray}
The Fourier transformed version of the hierarchy of equations
(\ref{level:n}) is:
\begin{eqnarray}
\hspace{-24mm}
\nonumber
        i\omega\bi{S} & = 0
\\
\nonumber
\hspace{-24mm}
        i\omega\bi{S}_{(1)} & = J\bi{S}_{(1)} \otimes \bi{S}
\\
\label{Fourier-level:n}
\hspace{-24mm}
i\omega\bi{S}_{(2)} & = J\bi{S}_{(2)} \otimes \bi{S} +
        (J_{(1)} - J)\bi{S}_{(2)} \otimes \bi{S}_{(1)} \\
\hspace{-22mm}
\nonumber
\ldots &
\\
\nonumber
\hspace{-24mm}
i\omega\bi{S}_{(n)} & = J\bi{S}_{(n)} \otimes \bi{S} +
        (J_{(1)} - J)\bi{S}_{(n)} \otimes \bi{S}_{(1)} + \ldots +
        (J_{(n-1)} - J_{(n-2)})\bi{S}_{(n)} \otimes \bi{S}_{(n-1)}
        \\ \nonumber
\hspace{-24mm}
i\omega\bi{S}_i & = J\bi{S}_i \otimes \bi{S} + (J_{(1)}
        - J)\bi{S}_i \otimes \bi{S}_{(1)} + \ldots + (J_{(n)} -
        J_{(n-1)})\bi{S}_{i} \otimes \bi{S}_{(n)}
\end{eqnarray}
where the symbol $\otimes$ denotes a vector product in the real space
and a convolution in the Fourier space.

        Although the explicit solution to the general case
(\ref{Fourier-level:n}) would be too cumbersome to write in a closed
form, it is easy to see that the first equation of
(\ref{Fourier-level:n}) will lead to $\bi{S}(\omega) \propto
\delta(\omega)$ (i.e., $\omega_{(0)} = 0$ or $\bi{S}$ is
constant). The second equation would introduce a new frequency in the
Fourier spectrum of $\bi{S}_{(1)}$: $\omega_{(1)} = JS$ (for instance,
this gives the spin precession frequency for each of spins in a
dimer\cite{dimer,dimer:field}). The third equation will give rise to
the additional frequency $\omega_{(2)} = (J_{(1)} - J)S_{(1)}$ in the
Fourier spectrum of $\bi{S}_{(2)}$ (as it happens for the individual
spins on the sites of a squashed equivalent neighbor model
\cite{isosceles, squashed}.) We are therefore led to state that the
solution for the individual spin $\bi{S}_i$ will contain
$\delta$-functions of various sums or differences of the frequencies
$JS$, $(J_{(1)} - J)S_{(1)}$, $(J_{(2)} - J_{(1)})S_{(2)}$, \ldots,
$(J_{(n)} - J_{(n-1)})S_{(n)}$ or, in other words, the Fourier
spectrum of $\bi{S}_i(t)$ will contain only these frequencies.


\section{Conclusions} \label{conclusions}

In this letter we have identified a class of integrable spin cluster
models and have presented a procedure for obtaining the exact
solutions for the time evolution of the individual spins, listing the
frequencies that appear in their Fourier spectra. We hope that these
results and solving procedures will prove useful to anyone interested
in easily identifying and solving specific classical Heisenberg spin
models relevant for describing the physical properties of magnetic
molecules.



\section*{References}
\begin{thebibliography}{99}
%
\bibitem{sriva1} Srivastava N, Kaufman C, M\"{u}ller G, Weber R,
        Thomas H, 1988 \ZP B {\bf 70} 251
%
\bibitem{Cr3tetra:2} Furukawa Y, Luban M, Borsa F, Johnston D C,
  Mahajan A V, Miller L L, Mentrup D, Schnack J  and Bino A 2000
  \PR B {\bf 61} 8635
%
 \bibitem{Fe3tetra:2} Bouwen A, Caneschi A, Gatteschi D, Goovaerts E,
  Schoemaker D, Sorace L and Stefan M 2001  \JPhCh {\bf 105}
  2658
%
\bibitem{Mn12} Cornia A, Sessoli R, Sorace L, Gatteschi D, Barra A L
  and Daiguebonne C 2002 \PRL {\bf 89} 257201
%
\bibitem{dimer} Luban M and Luscombe J
        1999 {\it Am.\ J.\ Phys.}  {\bf 67} 1161
%
\bibitem{square} Klemm R A and Luban M 2001 \PR B {\bf 64} 104424
%
\bibitem{isosceles} Ameduri M and Klemm R A 2002
        \PR B {\bf 66} 224404
%
\bibitem{dimer:field} Efremov D V and Klemm R A 2002 \PR B {\bf 66}
 174427
%
\bibitem{eq_neighbour} Klemm R A and Ameduri M 2002 \PR B {\bf 66}
 012403
%
\bibitem{squashed} Ameduri M and Klemm R A 2004 \JPA {\bf 37} 1095
%
\bibitem{knoxville} Haraldsen J T, Barnes T, Musfeldt J L 2004
  \textit{Neutron Scattering and Magnetic Observables for $S=1/2$
  Molecular Magnets} preprint cond-mat/0408253
%
\end{thebibliography}
\end{document}